\documentclass[sigconf]{acmart}
\AtBeginDocument{%
  }

\setcopyright{rightsretained}
\acmDOI{}
\acmConference[CHI '25 Workshop on Envisioning the Future of Interactive Health]{}{April 27th, 2025}{Yokohama, Japan}
\acmISBN{}
\copyrightyear{2025}
\acmYear{2025}

\begin{document}

\title[Facilitating Sensemaking on Contextualized Self-Tracking Data]{Facilitating Individuals' Sensemaking about Sedentary Behavior via Contextualized Data}

\author{Kefan Xu}
\affiliation{%
  \institution{Georgia Institute of Technology}
  \city{Atlanta}
  \state{Georgia}
  \country{USA}
  \postcode{30309}}
\email{kefanxu@gatech.edu}
\orcid{0000-0002-5492-8061}

\author{Rosa I. Arriaga}
\affiliation{%
  \institution{Georgia Institute of Technology}
  \city{Atlanta}
  \state{Georgia}
  \country{USA}
  \postcode{30309}}
\email{arriaga@cc.gatech.edu}
\orcid{0000-0002-8642-7245}


\begin{abstract}
The sedentary lifestyle increases individuals’ risks of developing chronic diseases. To support individuals to be more physically active, we propose a mobile system, MotionShift, that presents users with step count data alongside contextual information (e.g., location, weather, calendar events, etc.) and self-reported records. By implementing and deploying this system, we aim to understand how contextual information impacts individuals' sense-making on sensor-captured  data and how individuals leverage contextualized data to identify and reduce sedentary activities. The findings will advance the design of context-aware personal informatics systems, empowering users to derive actionable insights from sensor data while minimizing interpretation biases, ultimately promoting opportunities to be more physically active.

\end{abstract}




\maketitle

\section{Introduction}
Sedentary behaviors refer to activities that involve a very low energy expenditure (metabolic equivalents [MET] < 2.0) \cite{leslieEnvironmentalDeterminantsofPhysicalActivity}, mostly sitting and lying down \cite{onbehalfofsbrnterminologyconsensusprojectparticipantsSedentaryBehaviorResearch2017}. The sedentary lifestyle is one of the major causes of several chronic conditions (e.g., obesity \cite{silveiraSedentaryBehaviorPhysical2022}, cardiovascular disease \cite{lavieSedentaryBehaviorExercise2019, youngSedentaryBehaviorCardiovascular2016}, and diabetes \cite{silveiraSedentaryBehaviorPhysical2022}). While the CDC recommends 150 minutes of physical activity per week for maintaining physically active \cite{noauthor_physical_1999}, only 23\% of Americans meet this guideline \cite{stannerLeastFiveWeek2004, katzmarzykEpidemiologyPhysicalActivity2017}. More than 15\% of adults are physically inactive \cite{CDCNewsroom2016} with a notable disparity between races and ethnicities (i.e., 31.7\% of Hispanics, 30.3\% of non-Hispanic blacks, and 23.4\% of non-Hispanic whites). 

Despite the fact that few mobile tracking tools attempt to assist people in reducing sedentary behaviors \cite{intilleElicitingUserPreferences}, tracking sedentary behavior itself can be challenging due to its heterogeneity. It may be hard to distinguish sedentary activities from others with merely sensor-collected data. For instance, sedentary and stationary behaviors have similar data representations (e.g., low step counts), but a stationary behavior is not necessarily a sedentary behavior \cite{onbehalfofsbrnterminologyconsensusprojectparticipantsSedentaryBehaviorResearch2017}. Passive sitting (e.g., playing video games) and active sitting (e.g., working on a seated assembly line) both result in low step counts, but both are stationary behaviors with different levels of energy expenditure \cite{onbehalfofsbrnterminologyconsensusprojectparticipantsSedentaryBehaviorResearch2017}. Additionally, individuals perceive their energy consumption differently across various activities, making it challenging to identify sedentary behaviors. Both playing video games and working in front of a computer are screen activities \cite{onbehalfofsbrnterminologyconsensusprojectparticipantsSedentaryBehaviorResearch2017} but have different levels of mental energy consumption \cite{boolaniPhysicalActivityNot2021}, which can hardly be assessed by sensors. 

Moreover, even if one's sedentary time is identified, whether individuals can reduce it remains questionable and highly depends on context. Sedentary behaviors such as working in the office, having a meeting, and taking a rest after work may be different in nature in terms of how people can actually get rid of them. People can stand by their desks while working in the office to avoid sitting too long \cite{onbehalfofsbrnterminologyconsensusprojectparticipantsSedentaryBehaviorResearch2017}. But in most cases, people wouldn’t be able to stand up during an in-person meeting \cite{luoTimeBreakUnderstanding2018}. Prior study shows that there is an ambiguity in people’s understanding of sedentary behaviors (e.g., whether to attribute resting activities like sleep as sedentary behavior) \cite{kinnett-hopkinsInterpretationPhysicalActivity2019}. People may want to rest between different events and after work by lying, reclining, and sitting. Those activities are reflected as sedentary behaviors in the tracking data, but people need those activities to regain energy.

Reflecting on tracking data \cite{abowdChartingPresentFuture2000} has been found to help individuals make sense of their activity trends and patterns \cite{consolvoDesignRequirementsTechnologies2006, liUnderstandingMyData2011, paruthiFindingSweetSpot2018, xuUnderstandingEffectReflective2024, xuUnderstandingPeopleExperience2022}. However, many self-report tools are not designed to capture the context of sedentary behavior \cite{marinacFeasibilityUsingSenseCams2013}. When people are looking at their activity tracking data (e.g., step counts), it’s hard for them to recognize sedentary behaviors, nor can they tell if they can be more physically active at those times. Individuals in the previous study expressed their interests in reflecting on their active-inactive patterns to improve their physical activity level accordingly \cite{luoTimeBreakUnderstanding2018}. By identifying when they are most active, individuals are able to decide the right time to engage in physical activities \cite{alqahtaniSpatiotemporalContextualCues2022}. Though prior study shows that presenting individuals their physical activity data with contextual information helps improve the amount and quality of reflection \cite{alqahtaniSpatiotemporalContextualCues2022}, they also pointed out that not all those actions are feasible: some actions are hard to change, and some require long-term decisions \cite{alqahtaniSpatiotemporalContextualCues2022}. Even though the data captured by sensors can be considered objective, its meaning should be interpreted within the context where it’s collected \cite{pantzarLivingMetricsSelftracking2017}. 
\begin{figure*}
    \centering
    \includegraphics[width=1\linewidth]{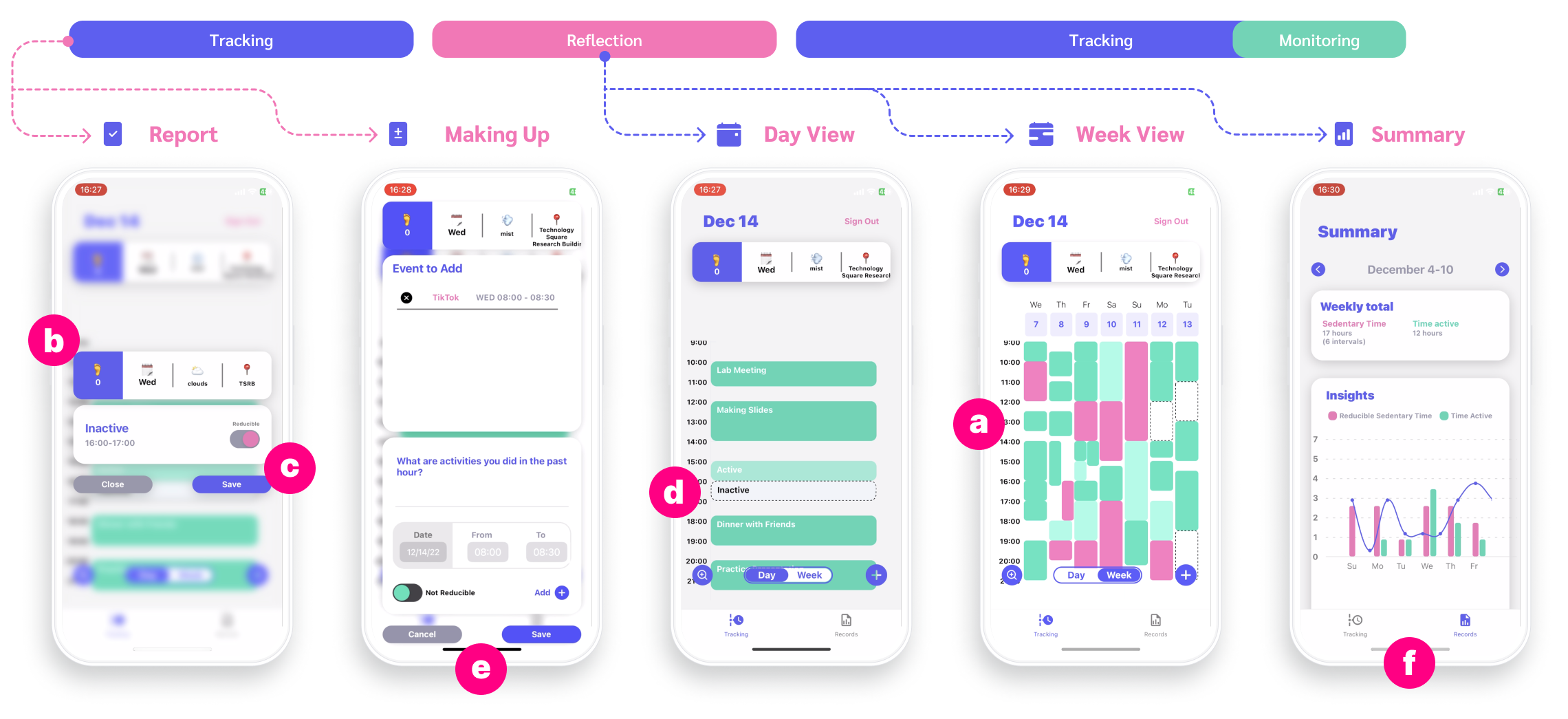}
    \caption{Proposed Design of MotionShift}
    \label{fig1}
\end{figure*}
Previous study highlights that missing contextual information makes it hard for people to interpret their physical activity tracking data \cite{tangHarnessingLongTerm2017} and leads to biased conclusions \cite{tangDefiningAdherenceMaking2018}. Also, to interpret sedentary behavior, there might be other data forms needed apart from contextual data. Ng et al. argue that data should be tracked and interpreted within its internal and external context \cite{ngUnderstandingSelfTrackedData2022a}. Prior research suggests combining subjective and objective measurements to provide information in the context of sedentary behavior \cite{janssenIssuesRelatedMeasuring2015}. Prince et al. also argue that future studies should complement objective measures with self-reported measures to assess sedentary behavior \cite{princeSingleMultiitemSelfassessment2018}. 

To sum up, previous studies suggest providing additional information (e.g., contextual information, subjective measurement, etc.) to assist people in investigating the nuance of sensor-collected data, especially in the context of tracking sedentary activities. In this proposed study, we raise the following research questions:

\begin{enumerate}
    \item How does contextual information impact the way individuals interpret sensor-captured  data (i.e., step counts)?
    \item How does contextualized data (e.g., weather, location, calendar events, etc.) support individuals in identifying and reducing sedentary periods in their daily lives?
\end{enumerate}

To address those questions, we proposed a mobile tracking system, MotionShift, as a probe to understand individuals' practice in tracking, identifying, and reducing their sedentary behaviors. By learning from users' experience using the system, we aim to examine individuals' sense-making on objective tracking data with other contextual information provided. Specifically, we want to understand how individuals interpret the nuances in visually similar data points. We hope the insights from this study can be generalizable to other fields (e.g., the use of sPGD in PTSD treatment \cite{ngProviderPerspectivesIntegrating2019a}) that treat sensor-collected objective data as the primary data form and highlight the needs of reflecting on it.  

\section{Related Works}

Combining contextual information with activity tracking data has long been studied in the HCI community. When presented with physical activity data and contextual information, individuals can make connections between them and be more aware of their activities \cite{liUsingContextReveal2012a}. Early literature points out that not tracking context can a pitfall for Q-selfers \cite{choeUnderstandingQuantifiedselfersPractices2014}. Echoing this insight, many personal informatics systems explore ways to present users their tracking data with contextual information. \citet{liangSleepExplorerVisualizationTool2016} visualized one's sleep data with other contextual factors to help users explore the correlation between them. \citet{bentleyHealthMashupsPresenting2013a} collected a variety of factors (e.g., step count, sleep, weight, etc.) from sensors and presented them to users with contextual data (e.g., weather, location, calendar, etc.). Findings from this study show that this combination leads to better self-understanding and behavioral change. Later research \cite{rajClinicalDataContext2019} treats clinical data (i.e., blood  glucose level) as the primary data type and leverages contextual information (e.g., mood, food, type of day) to help patients find trends.

Though sensor-collected tracking data is often considered the primary data type and subject to interpretation with contextual information, it may not be intuitive or reliable in many scenarios and can lead to biased conclusions. \citet{liangSleepExplorerVisualizationTool2016} demonstrated that tracking data may fall short in revealing sufficient insights when users maintain a consistent daily life pattern. Users may also abandon their tracking routine for a variety of reasons (e.g., seeing no values in tracking, etc.) \cite{epsteinReconsideringDeviceDrawer2016}, leaving gaps in their data. When users are away from their tracking devices, such missing data points may be misleading regarding their actual activity level \cite{bentleyHealthMashupsPresenting2013a}. Moreover, whether users can change their behavior using the knowledge they gained from tracking data remains uncertain. \citet{liangSleepExplorerVisualizationTool2016} pointed out the gap between individuals' intentions of behavior change and taking real actions, as the situation could be affected by factors that are not controllable. This uncertainty appears to be one of those factors that keep users from carrying out their physical activity plans \cite{xuUnderstandingPeopleExperience2022}. \citet{luoTimeBreakUnderstanding2018} conducted a study where they developed a break prompt system to encourage people to move more at work, showing that users' receptiveness for prompts highly depended on the context. Apart from those external barriers, people also face internal barriers that can impede them from conducting physical activities (e.g., lack of motivation \cite{astromSelfTrackingManagementPhysical2021}, physical or mental exhaustion \cite{xuUnderstandingPeopleExperience2022}, etc.). Lastly, people's perceptions towards their activity level can vary (e.g., different preferences for sleeping quality \cite{liangSleepExplorerVisualizationTool2016}), calling for a personalized way to assist individuals in interpreting their activity data.

\citet{pantzarLivingMetricsSelftracking2017} argue that the meaning of sensor-captured data is tied with contexts where it's collected, coining the concept of ``situated objectivity.'' Following this path, we propose to further investigate how contextual information could be better leveraged to assist individuals in making sense of their sensor-collected tracking data and reduce interpretation bias.

\section{Methodology}

In this study, we will develop a mobile app as a technology probe to understand how individuals can identify and reduce their sedentary behavior based on the step counts data with other contextual information provided. 

\subsection{System Design}

In response to \citet{liUsingContextReveal2012a}'s call for creating systems that allow people to easily draw connections between their physical activity and contextual information, the app, MotionShift, aims to assist individuals' sense-making with those data. The goal is to help users identify the sedentary time slots and try to turn those times into active times. It will feature a tracking-report-reflection flow that has been tested in the prior study \cite{xuUnderstandingEffectReflective2024}.

\subsubsection{Tracking}

The app will run in the background to collect users' step counts along with other contextual information (e.g., location, weather, etc.) on an hourly basis. We will implement an experience sampling method that's similar to Checkpoint-and-Remind \cite{changCombiningParticipatoryESM2020}. That's to say, if there are calendar events in the user's schedule, the app won't bother to get step count data from sensors, assuming those are events that do not add meaning to sensor-collected data (e.g., the time for having a meeting typically won't be considered as a time that the user can be more physically active). For the time slots that don't associate with any calendar events and have no step counts, the app will mark those time slots and wait for users to update what they have done afterward. Such a tracking process will result in different types of visualization, which we will elaborate on in the following section.

\subsubsection{Day/Week Views}

Users' step count data will be highlighted as the primary data type. The app will visualize step count data in calendar views (i.e., a day view and a week view). Depending on the threshold set by users regarding how many steps per hour count into active hours, the time blocks will be colored differently (Fig. \ref{fig1}a). To better illustrate potential patterns in sedentary behavior, the app will have a weekly view that presents the visualization based on step counts for the past 7 days. On both the day view and the week view, users can click into each color block, which represents an activity, to view further contextual information (Fig. \ref{fig1}b). Additionally, users can refine their records by manually indicating if the activity is a reducible sedentary activity or not (Fig. \ref{fig1}c). 

\subsubsection{Report}

As mentioned in section 3.1.1, there will be blank space on users' calendar view, meaning that there were no step counts nor calendar events during the time. For those time slots, users can choose to make up for their records. By clicking on those time slots (Fig. \ref{fig1}d), the user can provide more information regarding what they have done and determine if this time slot is a reducible sedentary time (Fig. \ref{fig1}e). Once again, contextual information will be provided on top of the screen to help users recall what happened. 

\subsubsection{Reflection}

Lastly, the app will offer users an overview of their data (Fig. \ref{fig1}f). On the summary page, the app will present users their weekly total sedentary time vs. active time. It will also summarize data into a chart for users to see trends in the data.

\subsection{User Study}

Once the app is ready to be deployed, we will recruit 16–20 participants who are willing to monitor their sedentary time and improve their physical activity level. We are particularly looking for participants (e.g., students, athletes, etc.) who are going through life changes so they will need to make sense of their data in different contexts (e.g., relocation, change of routines, etc.). We imagine participants' diverse lifestyles will add interesting data points regarding their interpretation of contextual information.

Participants will be using this app for 4 weeks and attending interviews at the beginning, middle point, and end of the study. The first interview aims to understand participants' lifestyles and their personal characteristics. The second interview will simply serve as a middle check-in point. And we will use the last interview to learn participants' experiences using the app, especially how they could make sense of their active-sedentary activities using step counts as the primary data type and interpret it along with other contextual information. We will quantitatively analyze the change of participants' physical tracking data (e.g., step counts) and qualitatively analyze the interviews.

\section{Expected Contribution}

This study will tackle the critical fact that sensor-collected data has been widely used as the primary data type for HCI research, while its meaning is subject to the situated context. We plan to investigate how individuals make sense of their step count data with contextual information annotated to identify the time when they can be more physically active. Potential insights from this study will shed light on how contextual information can be presented to individuals to reduce bias in interpreting sensor-collected data.

\bibliographystyle{ACM-Reference-Format}

\begin{thebibliography}{36}


\ifx \showCODEN    \undefined \def \showCODEN     #1{\unskip}     \fi
\ifx \showDOI      \undefined \def \showDOI       #1{#1}\fi
\ifx \showISBNx    \undefined \def \showISBNx     #1{\unskip}     \fi
\ifx \showISBNxiii \undefined \def \showISBNxiii  #1{\unskip}     \fi
\ifx \showISSN     \undefined \def \showISSN      #1{\unskip}     \fi
\ifx \showLCCN     \undefined \def \showLCCN      #1{\unskip}     \fi
\ifx \shownote     \undefined \def \shownote      #1{#1}          \fi
\ifx \showarticletitle \undefined \def \showarticletitle #1{#1}   \fi
\ifx \showURL      \undefined \def \showURL       {\relax}        \fi
\providecommand\bibfield[2]{#2}
\providecommand\bibinfo[2]{#2}
\providecommand\natexlab[1]{#1}
\providecommand\showeprint[2][]{arXiv:#2}

\bibitem[noa(1999)]%
        {noauthor_physical_1999}
 \bibinfo{year}{1999}\natexlab{}.
\newblock \bibinfo{title}{Physical {Inactivity} and {Cardiovascular} {Disease}}.
\newblock
\newblock
\urldef\tempurl%
\url{https://www.health.ny.gov/diseases/chronic/cvd.htm}
\showURL{%
\tempurl}


\bibitem[CDC(2016)]%
        {CDCNewsroom2016}
 \bibinfo{year}{2016}\natexlab{}.
\newblock \bibinfo{title}{{CDC} {Newsroom}}.
\newblock
\newblock
\urldef\tempurl%
\url{https://www.cdc.gov/media/releases/2020/0116-americas-inactivity.html}
\showURL{%
\tempurl}


\bibitem[Abowd and Mynatt(2000)]%
        {abowdChartingPresentFuture2000}
\bibfield{author}{\bibinfo{person}{Gregory~D. Abowd} {and} \bibinfo{person}{Elizabeth~D. Mynatt}.} \bibinfo{year}{2000}\natexlab{}.
\newblock \showarticletitle{Charting past, present, and future research in ubiquitous computing}.
\newblock \bibinfo{journal}{\emph{ACM Transactions on Computer-Human Interaction}} \bibinfo{volume}{7}, \bibinfo{number}{1} (\bibinfo{date}{March} \bibinfo{year}{2000}), \bibinfo{pages}{29--58}.
\newblock
\showISSN{1073-0516, 1557-7325}
\urldef\tempurl%
\url{https://doi.org/10.1145/344949.344988}
\showDOI{\tempurl}


\bibitem[Alqahtani et~al\mbox{.}(2022)]%
        {alqahtaniSpatiotemporalContextualCues2022}
\bibfield{author}{\bibinfo{person}{Deemah Alqahtani}, \bibinfo{person}{Caroline Jay}, {and} \bibinfo{person}{Markel Vigo}.} \bibinfo{year}{2022}\natexlab{}.
\newblock \showarticletitle{Spatio-temporal and contextual cues to support reflection in physical activity tracking}.
\newblock \bibinfo{journal}{\emph{International Journal of Human-Computer Studies}}  \bibinfo{volume}{165} (\bibinfo{date}{Sept.} \bibinfo{year}{2022}), \bibinfo{pages}{102865}.
\newblock
\showISSN{10715819}
\urldef\tempurl%
\url{https://doi.org/10.1016/j.ijhcs.2022.102865}
\showDOI{\tempurl}


\bibitem[Bentley et~al\mbox{.}(2013)]%
        {bentleyHealthMashupsPresenting2013a}
\bibfield{author}{\bibinfo{person}{Frank Bentley}, \bibinfo{person}{Konrad Tollmar}, \bibinfo{person}{Peter Stephenson}, \bibinfo{person}{Laura Levy}, \bibinfo{person}{Brian Jones}, \bibinfo{person}{Scott Robertson}, \bibinfo{person}{Ed Price}, \bibinfo{person}{Richard Catrambone}, {and} \bibinfo{person}{Jeff Wilson}.} \bibinfo{year}{2013}\natexlab{}.
\newblock \showarticletitle{Health {Mashups}: {Presenting} {Statistical} {Patterns} between {Wellbeing} {Data} and {Context} in {Natural} {Language} to {Promote} {Behavior} {Change}}.
\newblock \bibinfo{journal}{\emph{ACM Transactions on Computer-Human Interaction}} \bibinfo{volume}{20}, \bibinfo{number}{5} (\bibinfo{date}{Nov.} \bibinfo{year}{2013}), \bibinfo{pages}{1--27}.
\newblock
\showISSN{1073-0516, 1557-7325}
\urldef\tempurl%
\url{https://doi.org/10.1145/2503823}
\showDOI{\tempurl}


\bibitem[Boolani et~al\mbox{.}(2021)]%
        {boolaniPhysicalActivityNot2021}
\bibfield{author}{\bibinfo{person}{Ali Boolani}, \bibinfo{person}{Brandon Bahr}, \bibinfo{person}{Italia Milani}, \bibinfo{person}{Shane Caswell}, \bibinfo{person}{Nelson Cortes}, \bibinfo{person}{Matthew~Lee Smith}, {and} \bibinfo{person}{Joel Martin}.} \bibinfo{year}{2021}\natexlab{}.
\newblock \showarticletitle{Physical activity is not associated with feelings of mental energy and fatigue after being sedentary for 8 hours or more}.
\newblock \bibinfo{journal}{\emph{Mental Health and Physical Activity}}  \bibinfo{volume}{21} (\bibinfo{date}{Oct.} \bibinfo{year}{2021}), \bibinfo{pages}{100418}.
\newblock
\showISSN{17552966}
\urldef\tempurl%
\url{https://doi.org/10.1016/j.mhpa.2021.100418}
\showDOI{\tempurl}


\bibitem[Chang et~al\mbox{.}(2020)]%
        {changCombiningParticipatoryESM2020}
\bibfield{author}{\bibinfo{person}{Hsiu-Chi Chang}, \bibinfo{person}{Yung-Ju Chang}, \bibinfo{person}{Mark~W. Newman}, {and} \bibinfo{person}{Chih-Hsin Lin}.} \bibinfo{year}{2020}\natexlab{}.
\newblock \showarticletitle{Combining {Participatory} and {ESM}: {A} {Hybrid} {Approach} to {Collecting} {Annotated} {Mobility} {Data}}. In \bibinfo{booktitle}{\emph{Extended {Abstracts} of the 2020 {CHI} {Conference} on {Human} {Factors} in {Computing} {Systems}}}. \bibinfo{publisher}{ACM}, \bibinfo{address}{Honolulu HI USA}, \bibinfo{pages}{1--7}.
\newblock
\showISBNx{978-1-4503-6819-3}
\urldef\tempurl%
\url{https://doi.org/10.1145/3334480.3383066}
\showDOI{\tempurl}


\bibitem[Choe et~al\mbox{.}(2014)]%
        {choeUnderstandingQuantifiedselfersPractices2014}
\bibfield{author}{\bibinfo{person}{Eun~Kyoung Choe}, \bibinfo{person}{Nicole~B. Lee}, \bibinfo{person}{Bongshin Lee}, \bibinfo{person}{Wanda Pratt}, {and} \bibinfo{person}{Julie~A. Kientz}.} \bibinfo{year}{2014}\natexlab{}.
\newblock \showarticletitle{Understanding quantified-selfers' practices in collecting and exploring personal data}. In \bibinfo{booktitle}{\emph{Proceedings of the {SIGCHI} {Conference} on {Human} {Factors} in {Computing} {Systems}}}. \bibinfo{publisher}{ACM}, \bibinfo{address}{Toronto Ontario Canada}, \bibinfo{pages}{1143--1152}.
\newblock
\showISBNx{978-1-4503-2473-1}
\urldef\tempurl%
\url{https://doi.org/10.1145/2556288.2557372}
\showDOI{\tempurl}


\bibitem[Consolvo et~al\mbox{.}(2006)]%
        {consolvoDesignRequirementsTechnologies2006}
\bibfield{author}{\bibinfo{person}{Sunny Consolvo}, \bibinfo{person}{Katherine Everitt}, \bibinfo{person}{Ian Smith}, {and} \bibinfo{person}{James~A. Landay}.} \bibinfo{year}{2006}\natexlab{}.
\newblock \showarticletitle{Design requirements for technologies that encourage physical activity}. In \bibinfo{booktitle}{\emph{Proceedings of the {SIGCHI} {Conference} on {Human} {Factors} in {Computing} {Systems}}}. \bibinfo{publisher}{ACM}, \bibinfo{address}{Montréal Québec Canada}, \bibinfo{pages}{457--466}.
\newblock
\showISBNx{978-1-59593-372-0}
\urldef\tempurl%
\url{https://doi.org/10.1145/1124772.1124840}
\showDOI{\tempurl}


\bibitem[Epstein et~al\mbox{.}(2016)]%
        {epsteinReconsideringDeviceDrawer2016}
\bibfield{author}{\bibinfo{person}{Daniel~A. Epstein}, \bibinfo{person}{Jennifer~H. Kang}, \bibinfo{person}{Laura~R. Pina}, \bibinfo{person}{James Fogarty}, {and} \bibinfo{person}{Sean~A. Munson}.} \bibinfo{year}{2016}\natexlab{}.
\newblock \showarticletitle{Reconsidering the device in the drawer: lapses as a design opportunity in personal informatics}. In \bibinfo{booktitle}{\emph{Proceedings of the 2016 {ACM} {International} {Joint} {Conference} on {Pervasive} and {Ubiquitous} {Computing}}}. \bibinfo{publisher}{ACM}, \bibinfo{address}{Heidelberg Germany}, \bibinfo{pages}{829--840}.
\newblock
\showISBNx{978-1-4503-4461-6}
\urldef\tempurl%
\url{https://doi.org/10.1145/2971648.2971656}
\showDOI{\tempurl}


\bibitem[Intille et~al\mbox{.}({[n.\,d.]})]%
        {intilleElicitingUserPreferences}
\bibfield{author}{\bibinfo{person}{Stephen Intille}, \bibinfo{person}{Charles Kukla}, {and} \bibinfo{person}{Xiaoyi Ma}.} \bibinfo{year}{[n.\,d.]}\natexlab{}.
\newblock \showarticletitle{Eliciting {User} {Preferences} {Using} {Image}-{Based} {Experience} {Sampling} and {Reflection}}.
\newblock  (\bibinfo{year}{[n.\,d.]}), \bibinfo{pages}{2}.
\newblock


\bibitem[Janssen and Cliff(2015)]%
        {janssenIssuesRelatedMeasuring2015}
\bibfield{author}{\bibinfo{person}{Xanne Janssen} {and} \bibinfo{person}{Dylan~P. Cliff}.} \bibinfo{year}{2015}\natexlab{}.
\newblock \showarticletitle{Issues {Related} to {Measuring} and {Interpreting} {Objectively} {Measured} {Sedentary} {Behavior} {Data}}.
\newblock \bibinfo{journal}{\emph{Measurement in Physical Education and Exercise Science}} \bibinfo{volume}{19}, \bibinfo{number}{3} (\bibinfo{date}{July} \bibinfo{year}{2015}), \bibinfo{pages}{116--124}.
\newblock
\showISSN{1091-367X, 1532-7841}
\urldef\tempurl%
\url{https://doi.org/10.1080/1091367X.2015.1045908}
\showDOI{\tempurl}


\bibitem[Katzmarzyk et~al\mbox{.}(2017)]%
        {katzmarzykEpidemiologyPhysicalActivity2017}
\bibfield{author}{\bibinfo{person}{Peter~T. Katzmarzyk}, \bibinfo{person}{I-Min Lee}, \bibinfo{person}{Corby~K. Martin}, {and} \bibinfo{person}{Steven~N. Blair}.} \bibinfo{year}{2017}\natexlab{}.
\newblock \showarticletitle{Epidemiology of {Physical} {Activity} and {Exercise} {Training} in the {United} {States}}.
\newblock \bibinfo{journal}{\emph{Progress in Cardiovascular Diseases}} \bibinfo{volume}{60}, \bibinfo{number}{1} (\bibinfo{date}{July} \bibinfo{year}{2017}), \bibinfo{pages}{3--10}.
\newblock
\showISSN{00330620}
\urldef\tempurl%
\url{https://doi.org/10.1016/j.pcad.2017.01.004}
\showDOI{\tempurl}


\bibitem[Kinnett-Hopkins et~al\mbox{.}(2019)]%
        {kinnett-hopkinsInterpretationPhysicalActivity2019}
\bibfield{author}{\bibinfo{person}{Dominique Kinnett-Hopkins}, \bibinfo{person}{Yvonne Learmonth}, \bibinfo{person}{Elizabeth Hubbard}, \bibinfo{person}{Lara Pilutti}, \bibinfo{person}{Sarah Roberts}, \bibinfo{person}{Jason Fanning}, \bibinfo{person}{Thomas Wójcicki}, \bibinfo{person}{Edward McAuley}, {and} \bibinfo{person}{Robert Motl}.} \bibinfo{year}{2019}\natexlab{}.
\newblock \showarticletitle{The interpretation of physical activity, exercise, and sedentary behaviours by persons with multiple sclerosis}.
\newblock \bibinfo{journal}{\emph{Disability and Rehabilitation}} \bibinfo{volume}{41}, \bibinfo{number}{2} (\bibinfo{date}{Jan.} \bibinfo{year}{2019}), \bibinfo{pages}{166--171}.
\newblock
\showISSN{0963-8288, 1464-5165}
\urldef\tempurl%
\url{https://doi.org/10.1080/09638288.2017.1383519}
\showDOI{\tempurl}


\bibitem[Lavie et~al\mbox{.}(2019)]%
        {lavieSedentaryBehaviorExercise2019}
\bibfield{author}{\bibinfo{person}{Carl~J. Lavie}, \bibinfo{person}{Cemal Ozemek}, \bibinfo{person}{Salvatore Carbone}, \bibinfo{person}{Peter~T. Katzmarzyk}, {and} \bibinfo{person}{Steven~N. Blair}.} \bibinfo{year}{2019}\natexlab{}.
\newblock \showarticletitle{Sedentary {Behavior}, {Exercise}, and {Cardiovascular} {Health}}.
\newblock \bibinfo{journal}{\emph{Circulation Research}} \bibinfo{volume}{124}, \bibinfo{number}{5} (\bibinfo{date}{March} \bibinfo{year}{2019}), \bibinfo{pages}{799--815}.
\newblock
\showISSN{0009-7330, 1524-4571}
\urldef\tempurl%
\url{https://doi.org/10.1161/CIRCRESAHA.118.312669}
\showDOI{\tempurl}


\bibitem[Leslie et~al\mbox{.}({[n.\,d.]})]%
        {leslieEnvironmentalDeterminantsofPhysicalActivity}
\bibfield{author}{\bibinfo{person}{E Leslie}, \bibinfo{person}{J Salmon}, {and} \bibinfo{person}{M~J Fotheringham}.} \bibinfo{year}{[n.\,d.]}\natexlab{}.
\newblock \showarticletitle{Environmental {Determinantsof} {PhysicalActivity} and {SedentaryBehavior}}.
\newblock  (\bibinfo{year}{[n.\,d.]}), \bibinfo{pages}{7}.
\newblock


\bibitem[Li et~al\mbox{.}(2011)]%
        {liUnderstandingMyData2011}
\bibfield{author}{\bibinfo{person}{Ian Li}, \bibinfo{person}{Anind~K. Dey}, {and} \bibinfo{person}{Jodi Forlizzi}.} \bibinfo{year}{2011}\natexlab{}.
\newblock \showarticletitle{Understanding my data, myself: supporting self-reflection with ubicomp technologies}. In \bibinfo{booktitle}{\emph{Proceedings of the 13th international conference on {Ubiquitous} computing - {UbiComp} '11}}. \bibinfo{publisher}{ACM Press}, \bibinfo{address}{Beijing, China}, \bibinfo{pages}{405}.
\newblock
\showISBNx{978-1-4503-0630-0}
\urldef\tempurl%
\url{https://doi.org/10.1145/2030112.2030166}
\showDOI{\tempurl}


\bibitem[Li et~al\mbox{.}(2012)]%
        {liUsingContextReveal2012a}
\bibfield{author}{\bibinfo{person}{Ian Li}, \bibinfo{person}{Anind~K. Dey}, {and} \bibinfo{person}{Jodi Forlizzi}.} \bibinfo{year}{2012}\natexlab{}.
\newblock \showarticletitle{Using context to reveal factors that affect physical activity}.
\newblock \bibinfo{journal}{\emph{ACM Transactions on Computer-Human Interaction}} \bibinfo{volume}{19}, \bibinfo{number}{1} (\bibinfo{date}{March} \bibinfo{year}{2012}), \bibinfo{pages}{1--21}.
\newblock
\showISSN{1073-0516, 1557-7325}
\urldef\tempurl%
\url{https://doi.org/10.1145/2147783.2147790}
\showDOI{\tempurl}


\bibitem[Liang et~al\mbox{.}(2016)]%
        {liangSleepExplorerVisualizationTool2016}
\bibfield{author}{\bibinfo{person}{Zilu Liang}, \bibinfo{person}{Bernd Ploderer}, \bibinfo{person}{Wanyu Liu}, \bibinfo{person}{Yukiko Nagata}, \bibinfo{person}{James Bailey}, \bibinfo{person}{Lars Kulik}, {and} \bibinfo{person}{Yuxuan Li}.} \bibinfo{year}{2016}\natexlab{}.
\newblock \showarticletitle{{SleepExplorer}: a visualization tool to make sense of correlations between personal sleep data and contextual factors}.
\newblock \bibinfo{journal}{\emph{Personal and Ubiquitous Computing}} \bibinfo{volume}{20}, \bibinfo{number}{6} (\bibinfo{date}{Nov.} \bibinfo{year}{2016}), \bibinfo{pages}{985--1000}.
\newblock
\showISSN{1617-4909, 1617-4917}
\urldef\tempurl%
\url{https://doi.org/10.1007/s00779-016-0960-6}
\showDOI{\tempurl}


\bibitem[Luo et~al\mbox{.}(2018)]%
        {luoTimeBreakUnderstanding2018}
\bibfield{author}{\bibinfo{person}{Yuhan Luo}, \bibinfo{person}{Bongshin Lee}, \bibinfo{person}{Donghee~Yvette Wohn}, \bibinfo{person}{Amanda~L. Rebar}, \bibinfo{person}{David~E. Conroy}, {and} \bibinfo{person}{Eun~Kyoung Choe}.} \bibinfo{year}{2018}\natexlab{}.
\newblock \showarticletitle{Time for {Break}: {Understanding} {Information} {Workers}' {Sedentary} {Behavior} {Through} a {Break} {Prompting} {System}}. In \bibinfo{booktitle}{\emph{Proceedings of the 2018 {CHI} {Conference} on {Human} {Factors} in {Computing} {Systems}}}. \bibinfo{publisher}{ACM}, \bibinfo{address}{Montreal QC Canada}, \bibinfo{pages}{1--14}.
\newblock
\showISBNx{978-1-4503-5620-6}
\urldef\tempurl%
\url{https://doi.org/10.1145/3173574.3173701}
\showDOI{\tempurl}


\bibitem[Marinac et~al\mbox{.}(2013)]%
        {marinacFeasibilityUsingSenseCams2013}
\bibfield{author}{\bibinfo{person}{Catherine Marinac}, \bibinfo{person}{Gina Merchant}, \bibinfo{person}{Suneeta Godbole}, \bibinfo{person}{Jacqueline Chen}, \bibinfo{person}{Jacqueline Kerr}, \bibinfo{person}{Bronwyn Clark}, {and} \bibinfo{person}{Simon Marshall}.} \bibinfo{year}{2013}\natexlab{}.
\newblock \showarticletitle{The feasibility of using {SenseCams} to measure the type and context of daily sedentary behaviors}. In \bibinfo{booktitle}{\emph{Proceedings of the 4th {International} {SenseCam} \& {Pervasive} {Imaging} {Conference} on - {SenseCam} '13}}. \bibinfo{publisher}{ACM Press}, \bibinfo{address}{San Diego, California}, \bibinfo{pages}{42--49}.
\newblock
\showISBNx{978-1-4503-2247-8}
\urldef\tempurl%
\url{https://doi.org/10.1145/2526667.2526674}
\showDOI{\tempurl}


\bibitem[Ng et~al\mbox{.}(2019)]%
        {ngProviderPerspectivesIntegrating2019a}
\bibfield{author}{\bibinfo{person}{Ada Ng}, \bibinfo{person}{Rachel Kornfield}, \bibinfo{person}{Stephen~M. Schueller}, \bibinfo{person}{Alyson~K. Zalta}, \bibinfo{person}{Michael Brennan}, {and} \bibinfo{person}{Madhu Reddy}.} \bibinfo{year}{2019}\natexlab{}.
\newblock \showarticletitle{Provider {Perspectives} on {Integrating} {Sensor}-{Captured} {Patient}-{Generated} {Data} in {Mental} {Health} {Care}}.
\newblock \bibinfo{journal}{\emph{Proceedings of the ACM on Human-Computer Interaction}} \bibinfo{volume}{3}, \bibinfo{number}{CSCW} (\bibinfo{date}{Nov.} \bibinfo{year}{2019}), \bibinfo{pages}{1--25}.
\newblock
\showISSN{2573-0142}
\urldef\tempurl%
\url{https://doi.org/10.1145/3359217}
\showDOI{\tempurl}


\bibitem[Ng et~al\mbox{.}(2022)]%
        {ngUnderstandingSelfTrackedData2022a}
\bibfield{author}{\bibinfo{person}{Ada Ng}, \bibinfo{person}{Ashley~Marie Walker}, \bibinfo{person}{Laurie Wakschlag}, \bibinfo{person}{Nabil Alshurafa}, {and} \bibinfo{person}{Madhu Reddy}.} \bibinfo{year}{2022}\natexlab{}.
\newblock \showarticletitle{Understanding {Self}-{Tracked} {Data} from {Bounded} {Situational} {Contexts}}.
\newblock  (\bibinfo{year}{2022}), \bibinfo{pages}{1684--1697}.
\newblock
\urldef\tempurl%
\url{https://doi.org/10.1145/3532106.3533498}
\showDOI{\tempurl}


\bibitem[{on behalf of SBRN Terminology Consensus Project Participants} et~al\mbox{.}(2017)]%
        {onbehalfofsbrnterminologyconsensusprojectparticipantsSedentaryBehaviorResearch2017}
\bibfield{author}{\bibinfo{person}{{on behalf of SBRN Terminology Consensus Project Participants}}, \bibinfo{person}{Mark~S. Tremblay}, \bibinfo{person}{Salomé Aubert}, \bibinfo{person}{Joel~D. Barnes}, \bibinfo{person}{Travis~J. Saunders}, \bibinfo{person}{Valerie Carson}, \bibinfo{person}{Amy~E. Latimer-Cheung}, \bibinfo{person}{Sebastien~F.M. Chastin}, \bibinfo{person}{Teatske~M. Altenburg}, {and} \bibinfo{person}{Mai~J.M. Chinapaw}.} \bibinfo{year}{2017}\natexlab{}.
\newblock \showarticletitle{Sedentary {Behavior} {Research} {Network} ({SBRN}) – {Terminology} {Consensus} {Project} process and outcome}.
\newblock \bibinfo{journal}{\emph{International Journal of Behavioral Nutrition and Physical Activity}} \bibinfo{volume}{14}, \bibinfo{number}{1} (\bibinfo{date}{Dec.} \bibinfo{year}{2017}), \bibinfo{pages}{75}.
\newblock
\showISSN{1479-5868}
\urldef\tempurl%
\url{https://doi.org/10.1186/s12966-017-0525-8}
\showDOI{\tempurl}


\bibitem[Pantzar and Ruckenstein(2017)]%
        {pantzarLivingMetricsSelftracking2017}
\bibfield{author}{\bibinfo{person}{Mika Pantzar} {and} \bibinfo{person}{Minna Ruckenstein}.} \bibinfo{year}{2017}\natexlab{}.
\newblock \showarticletitle{Living the metrics: {Self}-tracking and situated objectivity}.
\newblock \bibinfo{journal}{\emph{DIGITAL HEALTH}}  \bibinfo{volume}{3} (\bibinfo{date}{Jan.} \bibinfo{year}{2017}), \bibinfo{pages}{2055207617712590}.
\newblock
\showISSN{2055-2076}
\urldef\tempurl%
\url{https://doi.org/10.1177/2055207617712590}
\showDOI{\tempurl}
\newblock
\shownote{Publisher: SAGE Publications Ltd}.


\bibitem[Paruthi et~al\mbox{.}(2018)]%
        {paruthiFindingSweetSpot2018}
\bibfield{author}{\bibinfo{person}{Gaurav Paruthi}, \bibinfo{person}{Shriti Raj}, \bibinfo{person}{Natalie Colabianchi}, \bibinfo{person}{Predrag Klasnja}, {and} \bibinfo{person}{Mark~W. Newman}.} \bibinfo{year}{2018}\natexlab{}.
\newblock \showarticletitle{Finding the {Sweet} {Spot}(s): {Understanding} {Context} to {Support} {Physical} {Activity} {Plans}}.
\newblock \bibinfo{journal}{\emph{Proceedings of the ACM on Interactive, Mobile, Wearable and Ubiquitous Technologies}} \bibinfo{volume}{2}, \bibinfo{number}{1} (\bibinfo{date}{March} \bibinfo{year}{2018}), \bibinfo{pages}{1--17}.
\newblock
\showISSN{2474-9567, 2474-9567}
\urldef\tempurl%
\url{https://doi.org/10.1145/3191761}
\showDOI{\tempurl}


\bibitem[Prince et~al\mbox{.}(2018)]%
        {princeSingleMultiitemSelfassessment2018}
\bibfield{author}{\bibinfo{person}{Stephanie~A. Prince}, \bibinfo{person}{Robert~D. Reid}, \bibinfo{person}{Jordan Bernick}, \bibinfo{person}{Anna~E. Clarke}, {and} \bibinfo{person}{Jennifer~L. Reed}.} \bibinfo{year}{2018}\natexlab{}.
\newblock \showarticletitle{Single versus multi-item self-assessment of sedentary behaviour: {A} comparison with objectively measured sedentary time in nurses}.
\newblock \bibinfo{journal}{\emph{Journal of Science and Medicine in Sport}} \bibinfo{volume}{21}, \bibinfo{number}{9} (\bibinfo{date}{Sept.} \bibinfo{year}{2018}), \bibinfo{pages}{925--929}.
\newblock
\showISSN{14402440}
\urldef\tempurl%
\url{https://doi.org/10.1016/j.jsams.2018.01.018}
\showDOI{\tempurl}


\bibitem[Raj et~al\mbox{.}(2019)]%
        {rajClinicalDataContext2019}
\bibfield{author}{\bibinfo{person}{Shriti Raj}, \bibinfo{person}{Joyce~M. Lee}, \bibinfo{person}{Ashley Garrity}, {and} \bibinfo{person}{Mark~W. Newman}.} \bibinfo{year}{2019}\natexlab{}.
\newblock \showarticletitle{Clinical {Data} in {Context}: {Towards} {Sensemaking} {Tools} for {Interpreting} {Personal} {Health} {Data}}.
\newblock \bibinfo{journal}{\emph{Proceedings of the ACM on Interactive, Mobile, Wearable and Ubiquitous Technologies}} \bibinfo{volume}{3}, \bibinfo{number}{1} (\bibinfo{date}{March} \bibinfo{year}{2019}), \bibinfo{pages}{1--20}.
\newblock
\showISSN{2474-9567}
\urldef\tempurl%
\url{https://doi.org/10.1145/3314409}
\showDOI{\tempurl}


\bibitem[Silveira et~al\mbox{.}(2022)]%
        {silveiraSedentaryBehaviorPhysical2022}
\bibfield{author}{\bibinfo{person}{Erika~Aparecida Silveira}, \bibinfo{person}{Carolina~Rodrigues Mendonça}, \bibinfo{person}{Felipe~Mendes Delpino}, \bibinfo{person}{Guilherme~Vinícius Elias~Souza}, \bibinfo{person}{Lorena Pereira~de Souza~Rosa}, \bibinfo{person}{Cesar de Oliveira}, {and} \bibinfo{person}{Matias Noll}.} \bibinfo{year}{2022}\natexlab{}.
\newblock \showarticletitle{Sedentary behavior, physical inactivity, abdominal obesity and obesity in adults and older adults: {A} systematic review and meta-analysis}.
\newblock \bibinfo{journal}{\emph{Clinical Nutrition ESPEN}}  \bibinfo{volume}{50} (\bibinfo{date}{Aug.} \bibinfo{year}{2022}), \bibinfo{pages}{63--73}.
\newblock
\showISSN{24054577}
\urldef\tempurl%
\url{https://doi.org/10.1016/j.clnesp.2022.06.001}
\showDOI{\tempurl}


\bibitem[Stanner(2004)]%
        {stannerLeastFiveWeek2004}
\bibfield{author}{\bibinfo{person}{S. Stanner}.} \bibinfo{year}{2004}\natexlab{}.
\newblock \showarticletitle{At {Least} {Five} a {Week}- a summary of the report from the {Chief} {Medical} {Officer} on physical activity}.
\newblock \bibinfo{journal}{\emph{Nutrition Bulletin}} \bibinfo{volume}{29}, \bibinfo{number}{4} (\bibinfo{date}{Dec.} \bibinfo{year}{2004}), \bibinfo{pages}{350--352}.
\newblock
\showISSN{1471-9827, 1467-3010}
\urldef\tempurl%
\url{https://doi.org/10.1111/j.1467-3010.2004.00455.x}
\showDOI{\tempurl}


\bibitem[Tang and Kay(2017)]%
        {tangHarnessingLongTerm2017}
\bibfield{author}{\bibinfo{person}{Lie~Ming Tang} {and} \bibinfo{person}{Judy Kay}.} \bibinfo{year}{2017}\natexlab{}.
\newblock \showarticletitle{Harnessing {Long} {Term} {Physical} {Activity} {Data}—{How} {Long}-term {Trackers} {Use} {Data} and {How} an {Adherence}-based {Interface} {Supports} {New} {Insights}}.
\newblock \bibinfo{journal}{\emph{Proceedings of the ACM on Interactive, Mobile, Wearable and Ubiquitous Technologies}} \bibinfo{volume}{1}, \bibinfo{number}{2} (\bibinfo{date}{June} \bibinfo{year}{2017}), \bibinfo{pages}{1--28}.
\newblock
\showISSN{2474-9567}
\urldef\tempurl%
\url{https://doi.org/10.1145/3090091}
\showDOI{\tempurl}


\bibitem[Tang et~al\mbox{.}(2018)]%
        {tangDefiningAdherenceMaking2018}
\bibfield{author}{\bibinfo{person}{Lie~Ming Tang}, \bibinfo{person}{Jochen Meyer}, \bibinfo{person}{Daniel~A. Epstein}, \bibinfo{person}{Kevin Bragg}, \bibinfo{person}{Lina Engelen}, \bibinfo{person}{Adrian Bauman}, {and} \bibinfo{person}{Judy Kay}.} \bibinfo{year}{2018}\natexlab{}.
\newblock \showarticletitle{Defining {Adherence}: {Making} {Sense} of {Physical} {Activity} {Tracker} {Data}}.
\newblock \bibinfo{journal}{\emph{Proceedings of the ACM on Interactive, Mobile, Wearable and Ubiquitous Technologies}} \bibinfo{volume}{2}, \bibinfo{number}{1} (\bibinfo{date}{March} \bibinfo{year}{2018}), \bibinfo{pages}{1--22}.
\newblock
\showISSN{2474-9567, 2474-9567}
\urldef\tempurl%
\url{https://doi.org/10.1145/3191769}
\showDOI{\tempurl}


\bibitem[Xu et~al\mbox{.}(2022)]%
        {xuUnderstandingPeopleExperience2022}
\bibfield{author}{\bibinfo{person}{Kefan Xu}, \bibinfo{person}{Xinghui Yan}, {and} \bibinfo{person}{Mark~W Newman}.} \bibinfo{year}{2022}\natexlab{}.
\newblock \showarticletitle{Understanding {People}’s {Experience} for {Physical} {Activity} {Planning} and {Exploring} the {Impact} of {Historical} {Records} on {Plan} {Creation} and {Execution}}. In \bibinfo{booktitle}{\emph{{CHI} {Conference} on {Human} {Factors} in {Computing} {Systems}}}. \bibinfo{publisher}{ACM}, \bibinfo{address}{New Orleans LA USA}, \bibinfo{pages}{1--15}.
\newblock
\showISBNx{978-1-4503-9157-3}
\urldef\tempurl%
\url{https://doi.org/10.1145/3491102.3501997}
\showDOI{\tempurl}


\bibitem[Xu et~al\mbox{.}(2024)]%
        {xuUnderstandingEffectReflective2024}
\bibfield{author}{\bibinfo{person}{Kefan Xu}, \bibinfo{person}{Xinghui~(Erica) Yan}, \bibinfo{person}{Myeonghan Ryu}, \bibinfo{person}{Mark~W Newman}, {and} \bibinfo{person}{Rosa~I. Arriaga}.} \bibinfo{year}{2024}\natexlab{}.
\newblock \showarticletitle{Understanding the {Effect} of {Reflective} {Iteration} on {Individuals}’ {Physical} {Activity} {Planning}}. In \bibinfo{booktitle}{\emph{Proceedings of the {CHI} {Conference} on {Human} {Factors} in {Computing} {Systems}}}. \bibinfo{publisher}{ACM}, \bibinfo{address}{Honolulu HI USA}, \bibinfo{pages}{1--17}.
\newblock
\showISBNx{9798400703300}
\urldef\tempurl%
\url{https://doi.org/10.1145/3613904.3641937}
\showDOI{\tempurl}


\bibitem[Young et~al\mbox{.}(2016)]%
        {youngSedentaryBehaviorCardiovascular2016}
\bibfield{author}{\bibinfo{person}{Deborah~Rohm Young}, \bibinfo{person}{Marie-France Hivert}, \bibinfo{person}{Sofiya Alhassan}, \bibinfo{person}{Sarah~M. Camhi}, \bibinfo{person}{Jane~F. Ferguson}, \bibinfo{person}{Peter~T. Katzmarzyk}, \bibinfo{person}{Cora~E. Lewis}, \bibinfo{person}{Neville Owen}, \bibinfo{person}{Cynthia~K. Perry}, \bibinfo{person}{Juned Siddique}, {and} \bibinfo{person}{Celina~M. Yong}.} \bibinfo{year}{2016}\natexlab{}.
\newblock \showarticletitle{Sedentary {Behavior} and {Cardiovascular} {Morbidity} and {Mortality}: {A} {Science} {Advisory} {From} the {American} {Heart} {Association}}.
\newblock \bibinfo{journal}{\emph{Circulation}} \bibinfo{volume}{134}, \bibinfo{number}{13} (\bibinfo{date}{Sept.} \bibinfo{year}{2016}).
\newblock
\showISSN{0009-7322, 1524-4539}
\urldef\tempurl%
\url{https://doi.org/10.1161/CIR.0000000000000440}
\showDOI{\tempurl}


\bibitem[Åström et~al\mbox{.}(2021)]%
        {astromSelfTrackingManagementPhysical2021}
\bibfield{author}{\bibinfo{person}{Fredrika Åström}, \bibinfo{person}{Jules Verkade}, \bibinfo{person}{Thijs de Kleijn}, {and} \bibinfo{person}{Armağan Karahanoğlu}.} \bibinfo{year}{2021}\natexlab{}.
\newblock \showarticletitle{Self-{Tracking} and {Management} of {Physical} {Activity} {Fluctuations}: {An} {Investigation} into {Seasons}}. In \bibinfo{booktitle}{\emph{Extended {Abstracts} of the 2021 {CHI} {Conference} on {Human} {Factors} in {Computing} {Systems}}}. \bibinfo{publisher}{ACM}, \bibinfo{address}{Yokohama Japan}, \bibinfo{pages}{1--7}.
\newblock
\showISBNx{978-1-4503-8095-9}
\urldef\tempurl%
\url{https://doi.org/10.1145/3411763.3451758}
\showDOI{\tempurl}


\end{thebibliography}



\end{document}